\documentclass[preprint, review, 3p, authoryear, times]{elsarticle}

\usepackage{todonotes}
\usepackage{subcaption} 

\usepackage{amsmath}  
\usepackage{capt-of}  
\usepackage{graphicx}
\usepackage{url}
\usepackage{color, soul}
\usepackage{bm}
\usepackage{amsmath} 
\usepackage{lineno}

\begin{document}

\title{
Automating Bibliometric Analysis with Sentence Transformers and Retrieval-Augmented Generation (RAG): A Pilot Study in Semantic and Contextual Search for Customized Literature Characterization for High-Impact Urban Research
}

\tnotetext[t1]{This manuscript has been authored by UT-Battelle, LLC, under contract DE-AC05-00OR22725 with the US Department of Energy (DOE). The US government retains and the publisher, by accepting the article for publication, acknowledges that the US government retains a nonexclusive, paid-up, irrevocable, worldwide license to publish or reproduce the published form of this manuscript, or allow others to do so, for US government purposes. DOE will provide public access to these results of federally sponsored research in accordance with the DOE Public Access Plan (http://energy.gov/downloads/doe-public-access-plan).}

\author[CSED]{Haowen Xu } \ead{xuh4@ornl.gov}
\author[UTK]{Xueping Li\corref{cor1}}\ead{xueping.li@utk.edu}
\author[UTK]{Jose Tupayachi}\ead{jtupayac@vols.utk.edu}
\author[BTRIC]{Jianming (Jamie) Lian}\ead{lianj@ornl.gov}
\author[CSED]{Femi Omitaomu}\ead{omitaomuoa@ornl.gov}

\cortext[cor1]{Corresponding author.}

\address[CSED]{Computational Sciences and Engineering Division, Oak Ridge National Laboratory, Oak Ridge, TN 37830, USA}
\address[UTK]{The Department of Industrial and Systems Engineering, The University of Tennessee, Knoxville, TN 37996, USA}
   
\address[BTRIC]{Building Technologies Research and Integration Center, Oak Ridge National Laboratory, Oak Ridge, TN 37830, USA}

\makeatletter
\newcommand{\printfnsymbol}[1]{%
  \textsuperscript{\@fnsymbol{#1}}%
}

\newcommand*{\MyIndent}{\hspace*{0.5cm}}%
\begin{abstract}
Bibliometric analysis is essential for understanding research trends, scope, and impact in urban science, especially in high-impact journals, such Nature Portfolios. However, traditional methods, relying on keyword searches and basic NLP techniques, often fail to uncover valuable insights not explicitly stated in article titles or keywords. These approaches are unable to 
perform semantic searches and contextual understanding, limiting their effectiveness in classifying topics and characterizing studies. In this paper, we address these limitations by leveraging Generative AI models, specifically transformers and Retrieval-Augmented Generation (RAG), to automate and enhance bibliometric analysis. We developed a technical workflow that integrates a vector database, Sentence Transformers, a Gaussian Mixture Model (GMM), Retrieval Agent, and Large Language Models (LLMs) to enable contextual search, topic ranking, and characterization of research using customized prompt templates. A pilot study analyzing 223 urban science-related articles published in Nature Communications over the past decade highlights the effectiveness of our approach in generating insightful summary statistics on the quality, scope, and characteristics of papers in high-impact journals. This study introduces a new paradigm for enhancing bibliometric analysis and knowledge retrieval in urban research, positioning an AI agent as a powerful tool for advancing research evaluation and understanding.
\end{abstract}




\maketitle

\section{Introduction}
Bibliometric analysis is a widely used method for evaluating and mapping research trends, impact, and scope across various scientific domains \citep{donthu2021conduct}. It provides quantitative insights by analyzing publication records, citations, and other scholarly outputs, helping researchers and policymakers understand the evolution of specific fields \citep{gan2022practical}. Over the past few decades, bibliometric analysis has evolved from basic citation counts and keyword frequency metrics to more sophisticated approaches, incorporating co-authorship networks, citation flows, and research topic clusters \citep{ramirez2019past}. These methods are particularly important in fields like urban science, where emerging topics such as smart cities require continuous monitoring to shape the direction of future research and innovation \citep{guo2019bibliometric}. Bibliometric analysis plays a key role in identifying influential works, emerging themes, and research gaps, thus guiding strategic decision-making in urban science and smart city development \citep{zhao2019mapping}.

However, traditional bibliometric methods face several limitations. Most rely heavily on keyword searches and basic text mining techniques, which depend on exact matches of terminologies and predefined keywords. These techniques often miss critical insights that are not explicitly captured in the titles or abstracts of research articles, thereby limiting the ability to fully understand and classify research topics \citep{romero2021learning}. Furthermore, traditional natural language processing (NLP) approaches, such as term frequency-inverse document frequency (TF-IDF) or simple word co-occurrence metrics, fail to capture the semantic meaning and contextual relationships between concepts\citep{safder2019bibliometric}. Although topic modeling methods like Latent Dirichlet Allocation (LDA) can offer significant benefits for bibliometric analysis by providing deeper insights into the relationships and structures within research literature \citep{chen2020structural}, they are primarily used for uncovering thematic structures and classifying article topics and are not designed for enabling semantic search or providing a contextual understanding of an article that involves deeper reasoning and interpretation. As a result, traditional bibliometric analysis often falls short in generating deeper insights that require a thorough review and interpretation of the article’s full content. Relying primarily on the analysis of titles, keywords, and standard metadata, limits the ability to provide a more customized and nuanced characterization of research based on the full textual data.

Recent advancements in generative AI models, such as large language models (LLMs), have opened new opportunities for enhancing research \citep{wang2024survey}. These models, including transformers and Retrieval-Augmented Generation (RAG) systems, excel at semantic understanding and contextual interpretation of complex texts, making them highly suitable for extracting valuable insights from research articles and technical manuals \citep{xu2024genai, tupayachi2024towards}. In the field of urban informatics, LLMs have been increasingly applied to analyze large volumes of text, uncovering patterns and trends that traditional methods would overlook \citep{liang2024gpt,xu2024leveraging}. In this paper, we propose a novel technical workflow for automating and enhancing bibliometric analysis by integrating Vector Databases, Sentence Transformers, Gaussian Mixture Models (GMM), Retrieval Agents, and an LLM. Our approach enables contextual search, topic ranking, and customized characterization of research articles, which we demonstrate through a pilot study analyzing 201 urban science-related articles published in Nature Communications over the past decade. This work addresses the limitations of traditional bibliometric methods, introducing a new paradigm for urban research analysis and knowledge retrieval through the development of AI agents with contextual understanding and reasoning capabilities. 

\section{Literature Review}
To overcome the limitations and knowledge gaps in traditional bibliometric analysis, recent studies have employed generative AI models, particularly transformer-based language models, to automate and enhance bibliometric methodologies.

\citet{fijavcko2024using} explores the application of generative AI in bibliometric analysis, focusing on 10 years of research abstracts from the European Resuscitation Congresses (ERC). Using ChatGPT-4, the study classified 2,491 abstracts into ERC guideline topics, with Basic Life Support and Adult Advanced Life Support being the most frequent. The research highlights the potential of large language models like ChatGPT-4 in categorizing and analyzing scientific literature and identifying trends. However, challenges included potential misclassification, the limited use of abstract titles rather than full-text, and heavy reliance on the model’s capabilities. These constraints highlight the challenges of automating bibliometric analysis in the absence of comprehensive datasets. However, the study effectively showcases the potential of AI to significantly improve bibliometric methodologies despite these limitations.

\citet{weng2022identification} introduces a methodology for detecting and visualizing key research topics using GPT-3 embeddings and the HDBSCAN clustering algorithm on 593 abstracts related to urban studies and machine learning. By clustering abstracts based on semantic similarity and extracting keywords using the Maximal Marginal Relevance (MMR) algorithm, the study provides an interactive tool for exploring abstract clusters and their associated topics. Challenges included optimizing clustering parameters and relying solely on abstracts, which may not fully represent the research. Some clusters contained outliers or minimal data, affecting accuracy. Despite these limitations, the study demonstrates the potential of transformer-based models in facilitating unsupervised bibliometric analysis, though refinement is needed.

Both articles emphasize the benefits of transformer-based and large language models for bibliometric analysis, while also addressing critical limitations such as data quality, optimization challenges, and input constraints when working with abstract-based datasets. To overcome these challenges, there is a need to harness recent advancements in sentence transformer models and RAG technologies. These innovations can enable the development of an AI agent capable of advanced contextual understanding of research articles, facilitating semantic search and providing tailored insights based on user-specific queries. This, in turn, can generate new bibliometric metrics, offering deeper and more comprehensive analysis.

\section{Methodology}
This section starts by outlining the design requirements for our proposed methods, then presents the conceptual workflow and its implementation, which combines Generative AI techniques with statistical models.

\begin{figure*}[hbt!]
    \includegraphics[width=1\textwidth]{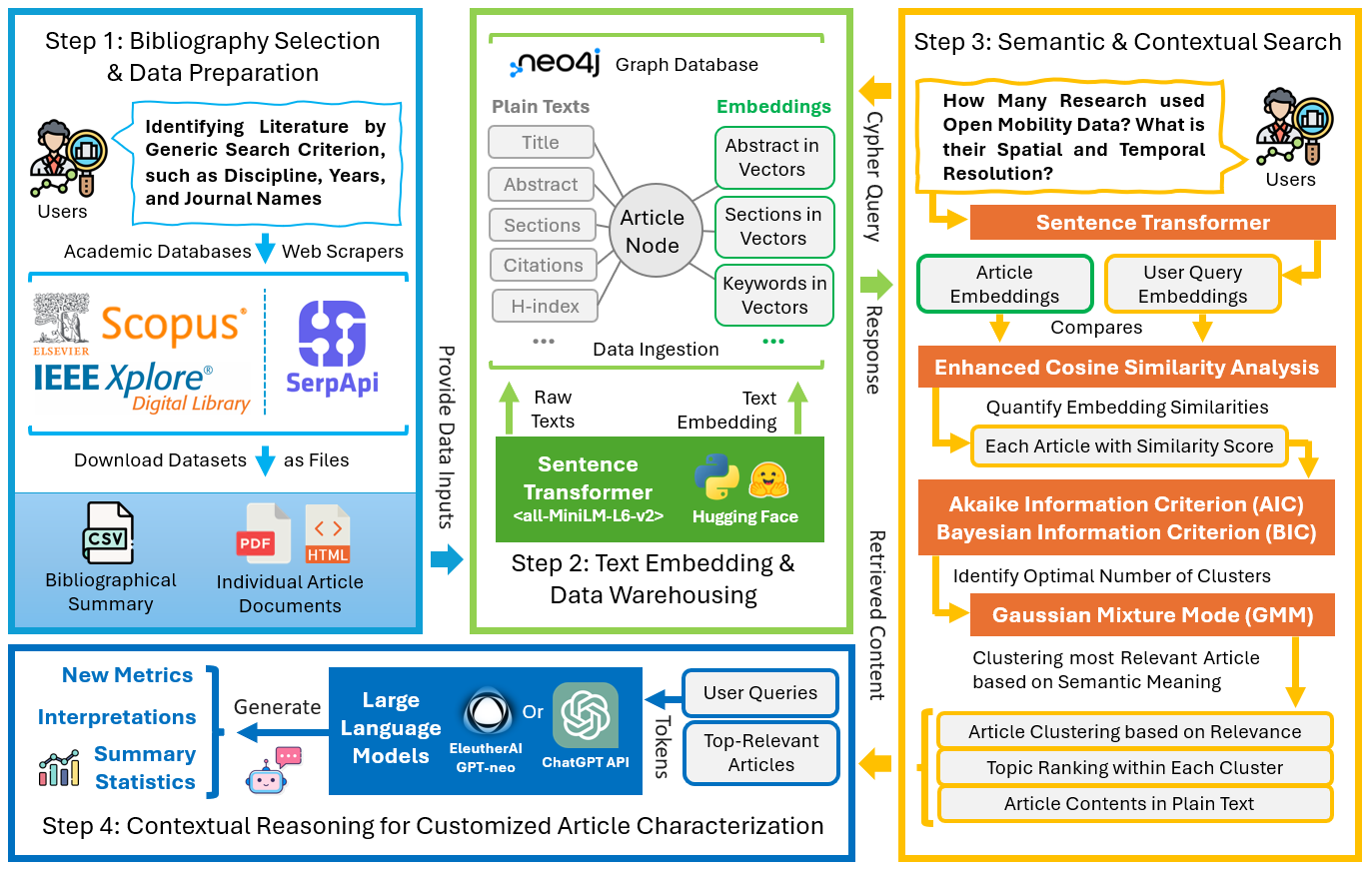}
    \caption{The overall design of the transformer and RAG-powered workflow.}
    \label{fig:workflow}
\end{figure*}

\subsection{Design Requirements}
Overall, we aim to develop an AI-agent styled tool that can interact with users, who are primarily researchers and college students, through human nature conversations, to get their inquire on the current-state of cutting edge research in a specific domain, such as smart city and urban science. Based on the inquiry, our workflow will automate a sequence of procedures that leverage the unique capabilities of sentence transformers and RAG techniques on a batch of selected literature filtered and downloaded from academics databases, such as Scopous, IEEE Xplore, and Web of Science. Aiming to shed lights on more advanced, intelligent, and automonous biblimetric analysis, our workflow aims to enable the following features: 
\begin{description}
    \item[Conversational Interaction:] A chatbot-style interface will be implemented, allowing users to ask questions through natural human conversations, without the need for pre-defined keywords or technical jargon. This feature will enable users to define search and filter criteria for subsets of bibliographic data (e.g., research articles, conference proceedings, technical reports, and manuals) that have been pre-selected and downloaded from popular academic databases. The search process will be guided by broad categories, such as domains, disciplines, and journals, to streamline access to relevant literature. 
    
    \item[Semantic and Contextual Search:] 
    Based on the user-defined inquiry, this process matches and retrieves relevant research documents or specific sections by analyzing the underlying meaning and contextual relationships between words, rather than relying solely on keyword matching. The use of sentence transformers and text embeddings, enables users to access information and knowledge based on conceptual relevance, rather than simple term frequency. This enhances the precision of literature filtering and facilitates deeper, more insightful knowledge discovery, which will plays important role as the retrieval agent within the RAG paradigm to benefit further analytics using Generative AI models.
 
    \item[Customized Literature Characterization:] Using the output literature from the semantic and contextual search as input, Generative Pre-trained Transformer (GPT) models will be employed for contextual understanding, reasoning, and interpretation. These GPT models will process user inquiries to generate customized characterizations and interpretations of the selected literature, providing deeper insights and creating more sophisticated metrics for advanced bibliometric analyses. This approach aims to enhance the overall understanding of research trends and offer tailored, in-depth evaluations of the literature.
\end{description}

As an example of our end-user capability, a user could ask the chatbot, powered by our method, a question like, ``What percentage of research published in Computers, Environment and Urban Systems over the past 5 years in the urban mobility sector uses traffic simulation-based methods versus crowd-sourced data-driven methods, and what are their spatial scales?'' The semantic and contextual search then filters and retrieves relevant articles based on the query, ranks them by relevance, and feeds them to the generative AI model. This enables advanced contextual understanding and reasoning to provide customized characterizations on individual research's simulation types and spatial scales, which involve information often not found in keywords or titles. These characterizations can be later used to generate summary statics and insights to facilitate more detailed trend analysis and thematic mapping.

\subsection{Workflow Design}
Our workflow consists of four key procedures, as depicted in Figure \ref{fig:workflow}. The workflow is later implemented in a Jupyter lab environment using Python-based libraries. Each procedure is detailed through the following following list. 

\subsubsection{Bibliography Selection and Data Preparation} 
In the first step, users select literature based on generic search criteria such as discipline, publication year, and journal name. Data is extracted from academic databases like Scopus, IEEE Xplore, and SerpApi using their respective web services and platforms, or through custom-built web scrapers, such as those powered by SerpAPI. The retrieved data includes bibliographic summaries in CSV format and individual articles in formats like PDF and HTML, which are then stored in a file-based system for further processing.

\subsubsection{Text Embedding and Data Warehousing} 
After retrieving the essential documents, a Python script powered by PyPDF2 is used to parse the bibliographic summaries, which include the list of downloaded articles along with supportive metadata (e.g., authors, year, source, citations, and h-index), as well as the PDF and HTML versions of the individual articles. This parsing process is designed to upload key textual information into a datastore, building the knowledge base for the proposed AI agent. Unlike traditional information and content management systems, our workflow utilizes a sentence transformer, specifically the all-MiniLM-L6-v2 model from Hugging Face, to generate text embeddings—vector representations that encode the semantic and contextual meaning of the text. Compared with traditional NLP methods, sentence transformers, with its unique self-attention mechanism, have superior advantages in capturing semantic meaning, enabling contextual understanding, handling synonyms, and long-range dependencies between words in a sentence. 
These embeddings facilitate more efficient semantic and contextual searches in later stages of the workflow. The text embeddings, along with essential metadata and article content, are uploaded into the datastore. We selected Neo4j, a graph database, as the datastore for this workflow due to its graph data model, which better represents the relationships between data entities stored as nodes in the database. In our project, individual articles are represented as nodes within Neo4j, with associated metadata, content, and text embeddings stored as properties of each node.

\subsubsection{Semantic and Contextual Search} 
In the third step, the workflow enables semantic and contextual searches within the literature stored in the knowledge base, leveraging the Neo4j database and sentence transformers. User queries, collected through a chatbot interface, serve as input for this advanced search. The core functionality compares the text embeddings of the user queries with those of the article contents. We employ an enhanced cosine similarity analysis, as described in Eq. \ref{eq:cosine_sim}, to calculate a similarity score ranging from 0 to 1, where 0 represents complete irrelevance and 1 represents high relevance. Our implementation extends the standard cosine similarity formula by using Python to chunk the original article content into sections and paragraphs, enabling more granular comparisons between the query and specific parts of the article. This process is applied to each article in the database, generating a similarity score based on semantic similarity with the user's query. At the contextual level, the framework evaluates the query’s context and intelligently selects embeddings from different sections of the articles to perform a targeted and accurate search.

\begin{equation}
\text{Similarity Score} = \frac{\mathbf{a} \cdot \mathbf{b}}{\|\mathbf{a}\| \|\mathbf{b}\|}
\label{eq:cosine_sim}  
\end{equation}
  
To draw the decision boundary based on a list of individual article's similarity score, we employed GMM to rank and cluster articles by their similarity score, which reflects their relevance. 
A GMM is a probabilistic model that represents a distribution of data as a mixture of multiple Gaussian (normal) distributions, each characterized by its own mean and variance, making it effective for modeling complex, multimodal datasets. We employed the Akaike Information Criterion (AIC) and Bayesian Information Criterion (BIC), alongside the elbow method, to determine the optimal number of clusters for the Gaussian Mixture Model (GMM) analysis. After the clustering analysis, the cluster with the highest average similarity scores implies it contains the most relevant articles, which are also further ranked based on its similarity score. 

Articles in the top-ranked clusters are subsequently fed into generative AI models, specifically GPT, to enable more in-depth analysis and interpretation. The semantic and contextual search within this workflow is a critical component of the RAG paradigm, allowing for further subsetting and refining of input information to ensure more accurate and relevant results. This process also helps prevent exceeding the token limits of the GPT model context window by optimizing the selection of input texts.

\subsubsection{Customized Article Characterization} 
The top-ranked clusters, containing the most relevant articles, are then imported into a GPT model as an external knowledge source to generate customized characterizations for each article. These tailored bibliographic characteristics serve as metrics, providing more detailed descriptions and classifications of the articles. This approach uncovers valuable insights into research trends, focusing on individual articles' topics, technologies, methods, and contributions.

We leverage the contextual reasoning capabilities of large language models to classify and justify findings based on the semantic meaning of sections and paragraphs within the articles, extracting useful information without relying on precise names or keywords. This process is guided by instructional prompting strategies, where we design engineered prompt templates and feed them into the GPT model along with the relevant article text segments (specific sections). These segments are further refined and filtered based on their content relevance to ensure accurate classification and extraction. 

At the technical level, we explored and tested the capabilities of two GPT models, including a local instance of EleutherAI/gpt-neo-1.3B models and the ChatGPT-3.5 Turbo API. Our experiments reveals that small models with on 1.3B parameters suffer from severe hallucination, and are unable to analyze large size of tokens. The ChatGPT-3.5 API demonstrates stable performance, particularly in its ability to process large text segments efficiently and produce reasoned characterizations.

\begin{figure*}[hbt!]
    \centering
    \begin{subfigure}[b]{1\textwidth}
        \centering
        \includegraphics[width=0.9\textwidth]{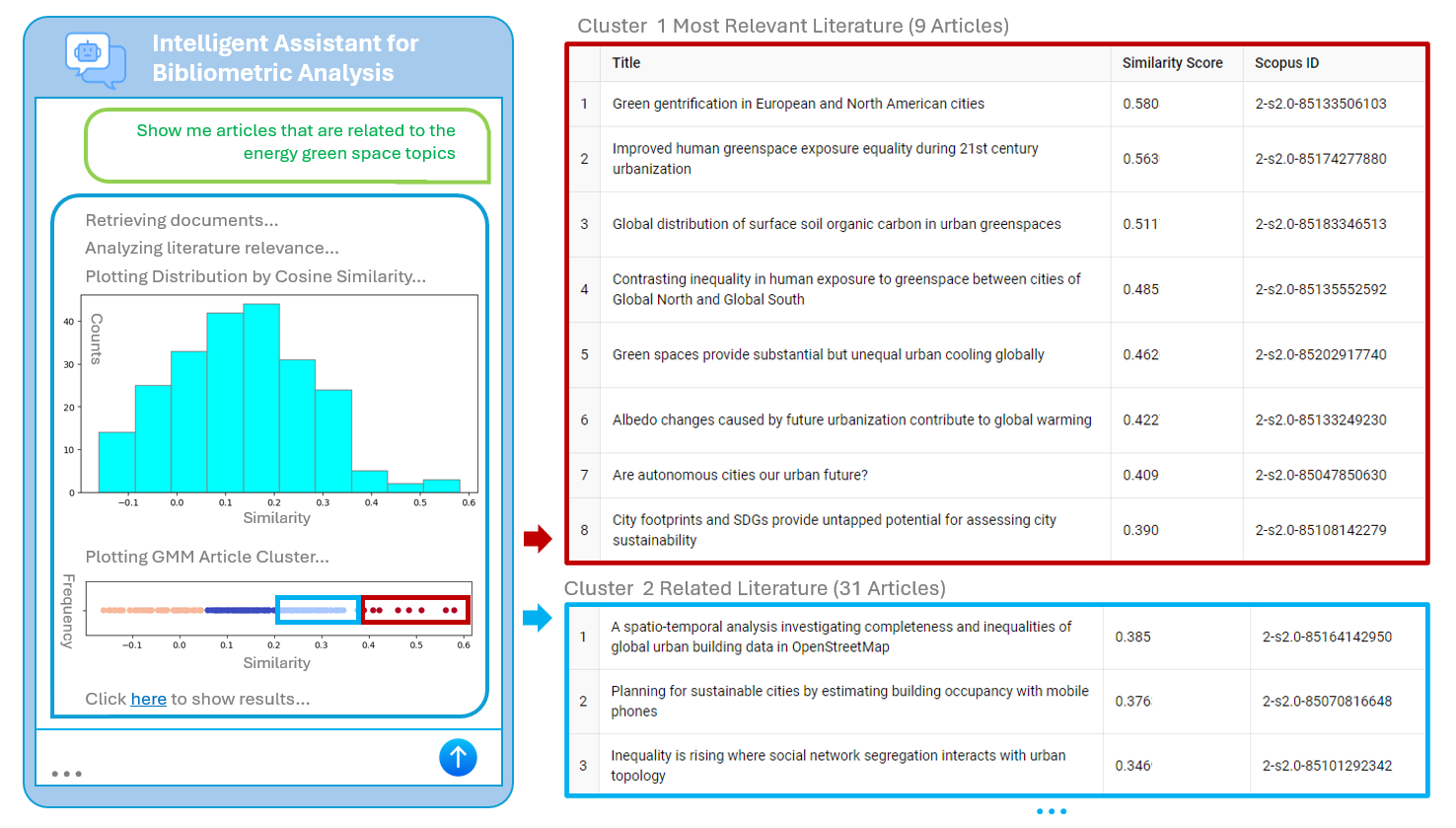}
        \caption{Semantic and contextual search based on user's query.}
        \label{fig:demo1}
    \end{subfigure}
    
    \vspace{1em} 
    
    \begin{subfigure}[b]{1\textwidth}
        \centering
        \includegraphics[width=0.9\textwidth]{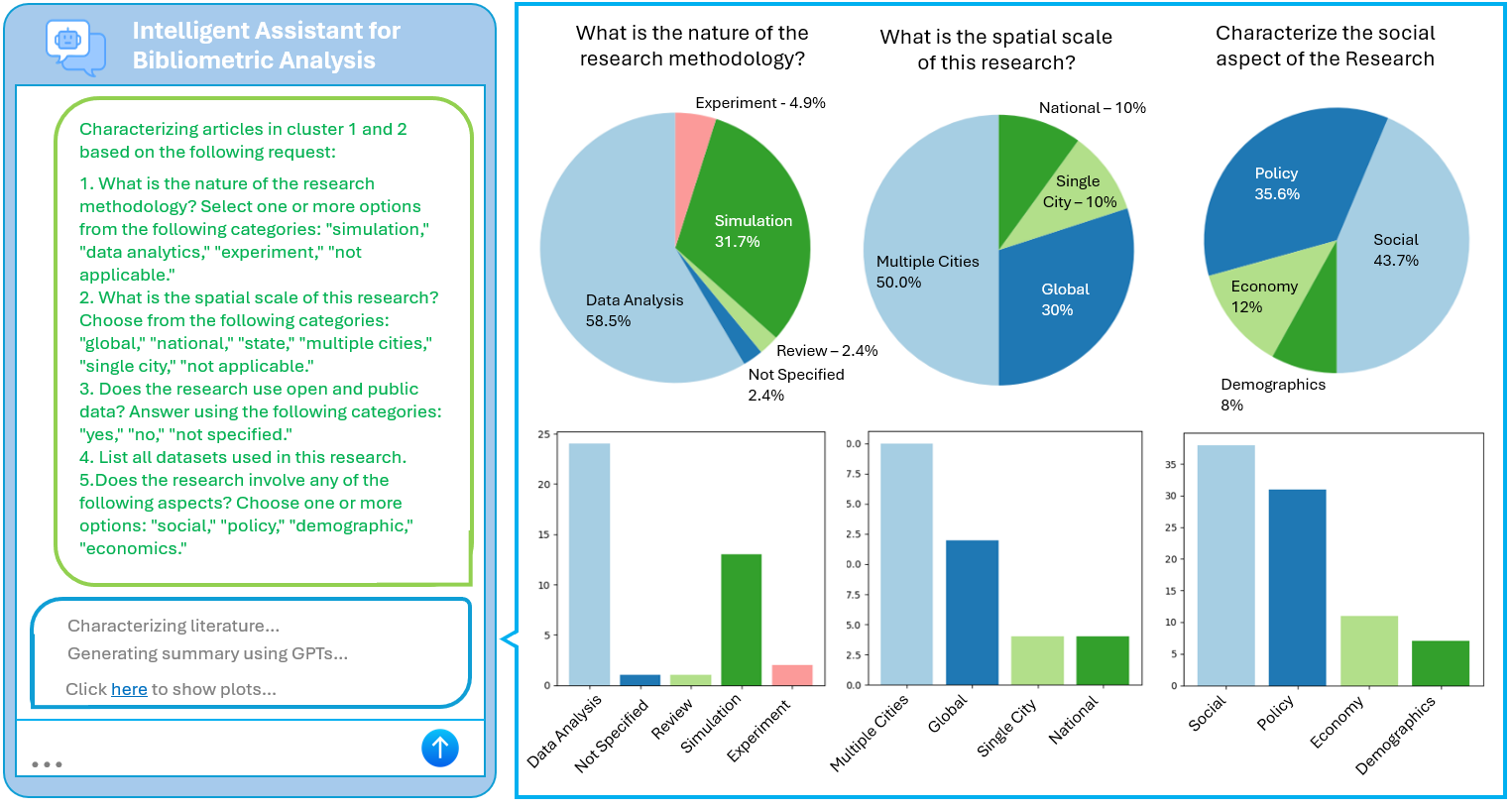}
        \caption{Customized Literature
            Characterization using GPT's reasoning capability.}
        \label{fig:demo2}
    \end{subfigure} 
    \caption{Demonstration of the workflow through two use case.}
\end{figure*}
  
\section{Pilot Study} 
This pilot study aims to demonstrate the feasibility and performance of our proposed methods. For this study, we compiled a dataset of 223 high-impact urban research articles published in Nature Communications, obtained through the following Scopus query: TITLE-ABS-KEY ( ``smart city'' OR ``urban'' OR ``urban management'' OR ``urban planning'' ) AND SRCTITLE ( ``Nature Communications'' ) AND PUBYEAR > 2013. We preprocessed the dataset by removing all intermediate versions labeled as ``Author Correction'' or ``Publisher Correction.'' The final dataset consists of a CSV file containing bibliometric summaries with all Scopus fields selected, along with 223 individual PDF documents of the actual articles.

\subsection{Use Case Demonstration}
Our first use case on semantic and contextual search is demonstrated in Figure \ref{fig:demo1}, where the user submits an inquiry to identify articles related to urban green space. The chatbot responds by visualizing a histogram of similarity scores for all articles and displaying the GMM clusters of the articles. Additionally, a link is provided to download a CSV file that ranks and clusters the articles based on their relevance to the user’s query.

Our second use case builds on the output articles from the first use case and demonstrates the capability to generate customized literature characterizations, creating new metrics for bibliometric analysis. Through the chatbot interface, users specify requests and instructions via prompts to guide the GPT model in generating tailored metrics. Examples of these prompts are illustrated in Figure \ref{fig:demo2}. Based on the prompts, the GPT processes the 67 retrieved articles and their critical content, leveraging its contextual reasoning ability to derive new literature characteristics, which can then be developed into metrics and summary statistics. Figure \ref{fig:demo2} also visualizes the responses to the user’s queries using pie charts and box plots. Users can submit additional questions and custom requests through the chatbot to extract information and develop unique metrics tailored to the needs of bibliometric analysis.

Our major contribution lies in the development of an autonomous AI agent designed to assist researchers in automating the characterization and information extraction of large volumes of literature, including datasets exceeding 1,000 articles. This system enables the generation of in-depth insights for bibliometric analysis, significantly enhancing the scalability and depth of literature review and research trend identification processes. By automating these tasks, the AI agent offers a powerful tool for efficiently managing and analyzing extensive collections of scholarly articles, ultimately facilitating more comprehensive and insightful bibliometric analyses.
 
\subsection{Limitation and Future Work}
Developed as a prototype for a more advanced knowledge base and management system, our workflow still faces a few limitations, as the following:
\begin{description}
    \item[Token Size Limitation:] The current implementation using the ChatGPT API has a maximum token size limitation and incurs service fees based on the number of tokens processed. This makes it less suitable for analyzing large volumes of literature.
    \item[Database Query Performance:] The current datastore implementation using the Neo4j database may encounter challenges in querying and managing large volumes of embedding data, as Neo4j is not optimized as a dedicated vector database.
    \item[Lack of Evaluation and Validation:] The GPT-generated literature characteristics are not currently evaluated by human experts, which introduces uncertainty regarding their accuracy and reliability.
\end{description}

As future work to address these limitations, we propose several experimental solutions. These include (a) deploying a local version of large language models, such as GPT-Neo, to minimize service fees for large-scale data analysis, (b) fine-tuning the GPT model to reduce unnecessary content and instructions sent to the model, thereby mitigating token size limitations, (c) transitioning our datastore implementation to dedicated vector databases, such as FAISS or Pinecone, to enhance latency and accuracy, and (d) developing a comprehensive strategy to evaluate the GPT's performance in analyzing and characterizing literature. Additionally, more advanced bibliometric analysis methods could be integrated into the current workflow to extend its analytical capabilities.

\section{Conclusion}
 In this paper, we have presented a novel workflow that integrates generative AI models and advanced analytical techniques through the RAG paradigm to address the limitations of traditional bibliometric analysis methods. By leveraging the contextual reasoning capabilities of large language models and enhanced semantic search techniques, our system offers a more nuanced and insightful analysis of research literature. This approach, demonstrated through the analysis of urban science-related articles, enables customized characterizations and generates new metrics for bibliometric analysis, providing deeper insights into research trends, methodologies, and contributions.

Our pilot study demonstrates the feasibility of this workflow, showcasing its ability to facilitate advanced semantic and contextual searches, cluster relevant articles, and produce tailored bibliographic insights through generative AI. However, the current implementation faces challenges, including token size limitations, database query performance issues, and the lack of expert evaluation for the AI-generated results.

To address these limitations, future work will explore the deployment of local language models, fine-tuning of GPT models to optimize token usage, and transitioning to vector databases like FAISS or Pinecone to improve performance. Additionally, we aim to establish a comprehensive validation framework involving human experts to ensure the accuracy and reliability of the generated bibliometric insights. As advancements in AI and bibliometric methodologies continue, our workflow has the potential to serve as a powerful and autonomous tool for researchers and policymakers seeking to analyze and interpret vast bodies of scientific literature more effectively.

\section{Acknowledgments}
This work was supported by the U.S. Department of Energy (U.S DOE),  Advanced Research Projects Agency–Energy (ARPA-E) under the project \#DE-AR0001780. We thank our collaborators from the University of Tennessee Knoxville.

\newpage

\small
\bibliographystyle{elsarticle-harv}
\bibliography{src}

\begin{thebibliography}{15}
\expandafter\ifx\csname natexlab\endcsname\relax\def\natexlab#1{#1}\fi
\providecommand{\url}[1]{\texttt{#1}}
\providecommand{\href}[2]{#2}
\providecommand{\path}[1]{#1}
\providecommand{\DOIprefix}{doi:}
\providecommand{\ArXivprefix}{arXiv:}
\providecommand{\URLprefix}{URL: }
\providecommand{\Pubmedprefix}{pmid:}
\providecommand{\doi}[1]{\href{http://dx.doi.org/#1}{\path{#1}}}
\providecommand{\Pubmed}[1]{\href{pmid:#1}{\path{#1}}}
\providecommand{\bibinfo}[2]{#2}
\ifx\xfnm\relax \def\xfnm[#1]{\unskip,\space#1}\fi
\bibitem[{Chen and Xie(2020)}]{chen2020structural}
\bibinfo{author}{Chen, X.}, \bibinfo{author}{Xie, H.}, \bibinfo{year}{2020}.
\newblock \bibinfo{title}{A structural topic modeling-based bibliometric study of sentiment analysis literature}.
\newblock \bibinfo{journal}{Cognitive Computation} \bibinfo{volume}{12}, \bibinfo{pages}{1097--1129}.
\bibitem[{Donthu et~al.(2021)Donthu, Kumar, Mukherjee, Pandey and Lim}]{donthu2021conduct}
\bibinfo{author}{Donthu, N.}, \bibinfo{author}{Kumar, S.}, \bibinfo{author}{Mukherjee, D.}, \bibinfo{author}{Pandey, N.}, \bibinfo{author}{Lim, W.M.}, \bibinfo{year}{2021}.
\newblock \bibinfo{title}{How to conduct a bibliometric analysis: An overview and guidelines}.
\newblock \bibinfo{journal}{Journal of business research} \bibinfo{volume}{133}, \bibinfo{pages}{285--296}.
\bibitem[{Fija{\v{c}}ko et~al.(2024)Fija{\v{c}}ko, Creber, Abella, Kocbek, Metli{\v{c}}ar, Greif and {\v{S}}tiglic}]{fijavcko2024using}
\bibinfo{author}{Fija{\v{c}}ko, N.}, \bibinfo{author}{Creber, R.M.}, \bibinfo{author}{Abella, B.S.}, \bibinfo{author}{Kocbek, P.}, \bibinfo{author}{Metli{\v{c}}ar, {\v{S}}.}, \bibinfo{author}{Greif, R.}, \bibinfo{author}{{\v{S}}tiglic, G.}, \bibinfo{year}{2024}.
\newblock \bibinfo{title}{Using generative artificial intelligence in bibliometric analysis: 10 years of research trends from the european resuscitation congresses}.
\newblock \bibinfo{journal}{Resuscitation Plus} \bibinfo{volume}{18}, \bibinfo{pages}{100584}.
\bibitem[{Gan et~al.(2022)Gan, Li, Robinson and Liu}]{gan2022practical}
\bibinfo{author}{Gan, Y.n.}, \bibinfo{author}{Li, D.d.}, \bibinfo{author}{Robinson, N.}, \bibinfo{author}{Liu, J.p.}, \bibinfo{year}{2022}.
\newblock \bibinfo{title}{Practical guidance on bibliometric analysis and mapping knowledge domains methodology--a summary}.
\newblock \bibinfo{journal}{European Journal of Integrative Medicine} \bibinfo{volume}{56}, \bibinfo{pages}{102203}.
\bibitem[{Guo et~al.(2019)Guo, Huang, Guo, Li, Guo and Nkeli}]{guo2019bibliometric}
\bibinfo{author}{Guo, Y.M.}, \bibinfo{author}{Huang, Z.L.}, \bibinfo{author}{Guo, J.}, \bibinfo{author}{Li, H.}, \bibinfo{author}{Guo, X.R.}, \bibinfo{author}{Nkeli, M.J.}, \bibinfo{year}{2019}.
\newblock \bibinfo{title}{Bibliometric analysis on smart cities research}.
\newblock \bibinfo{journal}{Sustainability} \bibinfo{volume}{11}, \bibinfo{pages}{3606}.
\bibitem[{Liang et~al.(2024)Liang, Zhao, Hou, Jin and Wu}]{liang2024gpt}
\bibinfo{author}{Liang, J.}, \bibinfo{author}{Zhao, A.}, \bibinfo{author}{Hou, S.}, \bibinfo{author}{Jin, F.}, \bibinfo{author}{Wu, H.}, \bibinfo{year}{2024}.
\newblock \bibinfo{title}{A gpt-enhanced framework on knowledge extraction and reuse for geographic analysis models in google earth engine}.
\newblock \bibinfo{journal}{International Journal of Digital Earth} \bibinfo{volume}{17}, \bibinfo{pages}{2398063}.
\bibitem[{Ram{\'\i}rez et~al.(2019)Ram{\'\i}rez, S{\'a}nchez-Ca{\~n}izares and Fuentes-Garc{\'\i}a}]{ramirez2019past}
\bibinfo{author}{Ram{\'\i}rez, L.J.C.}, \bibinfo{author}{S{\'a}nchez-Ca{\~n}izares, S.M.}, \bibinfo{author}{Fuentes-Garc{\'\i}a, F.J.}, \bibinfo{year}{2019}.
\newblock \bibinfo{title}{Past themes and tracking research trends in entrepreneurship: A co-word, cites and usage count analysis}.
\newblock \bibinfo{journal}{Sustainability} \bibinfo{volume}{11}, \bibinfo{pages}{3121}.
\bibitem[{Romero-Silva and De~Leeuw(2021)}]{romero2021learning}
\bibinfo{author}{Romero-Silva, R.}, \bibinfo{author}{De~Leeuw, S.}, \bibinfo{year}{2021}.
\newblock \bibinfo{title}{Learning from the past to shape the future: A comprehensive text mining analysis of or/ms reviews}.
\newblock \bibinfo{journal}{Omega} \bibinfo{volume}{100}, \bibinfo{pages}{102388}.
\bibitem[{Safder and Hassan(2019)}]{safder2019bibliometric}
\bibinfo{author}{Safder, I.}, \bibinfo{author}{Hassan, S.U.}, \bibinfo{year}{2019}.
\newblock \bibinfo{title}{Bibliometric-enhanced information retrieval: a novel deep feature engineering approach for algorithm searching from full-text publications}.
\newblock \bibinfo{journal}{Scientometrics} \bibinfo{volume}{119}, \bibinfo{pages}{257--277}.
\bibitem[{Tupayachi et~al.(2024)Tupayachi, Xu, Omitaomu, Camur, Sharmin and Li}]{tupayachi2024towards}
\bibinfo{author}{Tupayachi, J.}, \bibinfo{author}{Xu, H.}, \bibinfo{author}{Omitaomu, O.A.}, \bibinfo{author}{Camur, M.C.}, \bibinfo{author}{Sharmin, A.}, \bibinfo{author}{Li, X.}, \bibinfo{year}{2024}.
\newblock \bibinfo{title}{Towards next-generation urban decision support systems through ai-powered construction of scientific ontology using large language models—a case in optimizing intermodal freight transportation}.
\newblock \bibinfo{journal}{Smart Cities} \bibinfo{volume}{7}, \bibinfo{pages}{2392--2421}.
\bibitem[{Wang et~al.(2024)Wang, Ma, Feng, Zhang, Yang, Zhang, Chen, Tang, Chen, Lin et~al.}]{wang2024survey}
\bibinfo{author}{Wang, L.}, \bibinfo{author}{Ma, C.}, \bibinfo{author}{Feng, X.}, \bibinfo{author}{Zhang, Z.}, \bibinfo{author}{Yang, H.}, \bibinfo{author}{Zhang, J.}, \bibinfo{author}{Chen, Z.}, \bibinfo{author}{Tang, J.}, \bibinfo{author}{Chen, X.}, \bibinfo{author}{Lin, Y.}, et~al., \bibinfo{year}{2024}.
\newblock \bibinfo{title}{A survey on large language model based autonomous agents}.
\newblock \bibinfo{journal}{Frontiers of Computer Science} \bibinfo{volume}{18}, \bibinfo{pages}{186345}.
\bibitem[{Weng et~al.(2022)Weng, Wu and Dyer}]{weng2022identification}
\bibinfo{author}{Weng, M.H.}, \bibinfo{author}{Wu, S.}, \bibinfo{author}{Dyer, M.}, \bibinfo{year}{2022}.
\newblock \bibinfo{title}{Identification and visualization of key topics in scientific publications with transformer-based language models and document clustering methods}.
\newblock \bibinfo{journal}{Applied Sciences} \bibinfo{volume}{12}, \bibinfo{pages}{11220}.
\bibitem[{Xu et~al.(2024a)Xu, Omitaomu, Sabri, Zlatanova, Li and Song}]{xu2024leveraging}
\bibinfo{author}{Xu, H.}, \bibinfo{author}{Omitaomu, F.}, \bibinfo{author}{Sabri, S.}, \bibinfo{author}{Zlatanova, S.}, \bibinfo{author}{Li, X.}, \bibinfo{author}{Song, Y.}, \bibinfo{year}{2024}a.
\newblock \bibinfo{title}{Leveraging generative ai for urban digital twins: A scoping review on the autonomous generation of urban data, scenarios, designs, and 3d city models for smart city advancement}.
\newblock \bibinfo{journal}{arXiv preprint arXiv:2405.19464} \URLprefix \url{https://arxiv.org/abs/2405.19464}, \DOIprefix\doi{10.48550/arXiv.2405.19464}, \href{http://arxiv.org/abs/2405.19464}{{\tt arXiv:2405.19464}}. \bibinfo{note}{computer Science > Artificial Intelligence}.
\bibitem[{Xu et~al.(2024b)Xu, Yuan, Zhou, Xu, Li, Ye et~al.}]{xu2024genai}
\bibinfo{author}{Xu, H.}, \bibinfo{author}{Yuan, J.}, \bibinfo{author}{Zhou, A.}, \bibinfo{author}{Xu, G.}, \bibinfo{author}{Li, W.}, \bibinfo{author}{Ye, X.}, et~al., \bibinfo{year}{2024}b.
\newblock \bibinfo{title}{Genai-powered multi-agent paradigm for smart urban mobility: Opportunities and challenges for integrating large language models (llms) and retrieval-augmented generation (rag) with intelligent transportation systems}.
\newblock \bibinfo{journal}{arXiv preprint arXiv:2409.00494} .
\bibitem[{Zhao et~al.(2019)Zhao, Tang and Zou}]{zhao2019mapping}
\bibinfo{author}{Zhao, L.}, \bibinfo{author}{Tang, Z.y.}, \bibinfo{author}{Zou, X.}, \bibinfo{year}{2019}.
\newblock \bibinfo{title}{Mapping the knowledge domain of smart-city research: A bibliometric and scientometric analysis}.
\newblock \bibinfo{journal}{Sustainability} \bibinfo{volume}{11}, \bibinfo{pages}{6648}.

\end{thebibliography}

\end{document}